\documentstyle[preprint,aps]{revtex}
\begin{document}
\title{Observation of a new excitation in bcc $^{4}$He by inelastic neutron
scattering}
\author{T. Markovich$^{a}$, E. Polturak$^{a}$, J. Bossy$^{b}$, and E. Farhi$^{c}$}
\address{(a)Physics Department, Technion - IIT, Haifa, Israel 3200, \\
(b) CNRS-CRTBT, BP166, 38042 Grenoble Cedex 9,France, \\
(c) Institut Laue Langevin, BP 156, 38042 Grenoble Cedex 9,France.}
\date{\today}
\maketitle

\begin{abstract}
We report neutron scattering measurements of the phonons in bcc solid $^{4}$%
He. In general, only 3 accoustic phonon branches should exist in a
monoatomic cubic crystal. In addition to these phonon branches, we found a
new ''optic-like'' mode along the [110] direction. One possible
interpretation of this new mode is in terms of localized excitations unique
to a quantum solid.
\end{abstract}

The main attribute of a quantum solid is the large zero point motion of the
atoms. The vibrating atoms thus encounter the repulsive part of the
interatomic potential, leading to strong short range correlations.
Theoretical description of the solid is consequently a tour de force study
of anharmonic effects. The self consistent phonon (SCP) theory which emerged
over the years, treats these short range correlations using Jastrow-type
wavefunctions in the description of the ground state\cite
{nosanow,guyer1,horner,koehler,varma,glyde1}, in addition to the inclusion
of anharmonic terms in the potential. Phonon dispersion curves calculated in
the SCP theory are in good agreement with neutron scattering experiments on
solid $^{4}$He \cite{minkiewicz,osgood}, except for the T$_{1}$[110] branch
of bcc $^{4}$He, which experimentally has a much lower energy. In addition,
the SCP theory was able to explain several observations unique to quantum
solids, such as multiphonon scattering which dominates the scattering
already at intermediate momentum transfer Q, and phonon interference which
gives rise to oscillations of the effective{\em \ }Debye Waller factor. Our
motivation to do new experiments came following a recently proposed model 
\cite{gov}. This model aims to explain the discrepancy regarding the T$_{1}$%
[110] branch of bcc $^{4}$He through an approach in which the zero point
motion is treated separately from the usual harmonic degrees of freedom.
Indeed, the T$_{1}$[110] mode predicted by the new model agrees with
experiment, while leaving the other phonons unaffected. In addition, the
model\ predicts the appearance of a new dispersionless excitation branch at
an energy twice that of the T$_{1}$[110] phonon at the zone boundary ($\sim $%
1.2meV). Such an ''optic-like'' mode was not reported\cite{minkiewicz,osgood}%
.

In order to check this prediction, we performed inelastic neutron scattering
experiments on the bcc phase of $^{4}$He using the IN-12 spectrometer at the
ILL. $^{4}$He single crystals were grown and oriented in the beam. Since the
maximal temperature width of the bcc phase is less than 50 mK, particular
care was devoted to temperature control, resulting in a $\pm $0.1mK
stability. In addition, the growth cell contained a capacitive pressure
gauge allowing us to precisely control the location on the phase diagram.
Out of the 5 crystals grown, 4 could be oriented in the \{11$\overline{1}$\}
scattering plane, and one in the \{100\}. From the observed Bragg peaks, the
cell contained only one large single crystal. The energy of the incident
beam could be varied in the range of 2.3 to 14 meV. Constant-Q scans were
usually done at fixed k$_{F}$. In the high resolution scans we used a cooled
Be filter to remove $\lambda /2$ contamination. The highest instrumental
resolution used was 0.07 meV (FWHM) with k$_{F}$=1.15 1/\AA . Data were
typically taken around the (1,1,0) point.

In addition to well defined single phonon groups and broad multiphonon
features, the measured spectra also show a new, well defined feature. In{\em %
\ }Fig. 1, we show an intensity map displaying longitudinal scans in which
the new feature appears as a weakly dispersive ''optic-like'' branch,
together with the L[110] phonons. The resolution in Fig. 1 is not sufficient
to display details of the spectra, which are seen more clearly in Fig. 2.%
{\footnotesize \ } The new feature was observed in all 4 crystals during
scans along the [110] direction. This feature disappeared once the crystal
was molten and the cell contained only liquid. It is therefore not connected
with scattering from the liquid or from the walls of the cell . Moreover, we
performed intentional experiments in the mixed phase region, with both hcp
and bcc crystals in the cell. We observed no correlation between the phonon
peaks of the hcp phase and the new feature. We conclude therefore that the
new feature is a property of the bcc solid.

The average energy (over all 4 crystals) of this new mode at the zone origin
is 1.23 meV, almost exactly twice the energy of the zone boundary T$_{1}$
phonon (0.59$\pm $0.01 meV). This ratio is constant over the whole
temperature range where the bcc phase exists. The best location to observe
this mode was at a reduced wavevector $q$=0, where it was not masked by any
of the other phonon branches. However, as Fig. 1 shows, it is possible to
resolve this mode up to $q$=0.2 in units of $2\pi /a$ (r.l.u.). At higher $q$%
, the phonon groups are too broad to identify this feature. In Fig.3, we
show the dispersion relations measured along the [110] direction. The lines
in the figure are from the SCP theory \cite{glyde2} and from the new model
\cite{gov}. Our data for the three phonon branches is in excellent agreement
with previous experiments\cite{minkiewicz,osgood}. The fact that the T$_{1}$
branch is lower by a factor of $\sim $ 2 compared with the SCP calculation
was already evident from the previous experiments \cite{minkiewicz,osgood},
and is confirmed in the present work. In addition, Fig. 3 shows the data for
the new feature, which shows some dispersion. It is important to ask whether
the new feature is a distinct mode, rather than a product of anharmonic
effects described by the SCP theory\cite
{glydebook,choquard,boccara,koehler2,horner2}, namely phonon interference
and multiphonon scattering. First, phonon interference gives rise to
scattering intensity which is an odd function of $q$, namely it vanishes at
the zone origin\cite{glydebook}. The fact that we see this feature clearly
at the zone origin excludes the possibility that it is due to interference
effects. Regarding multiphonon effects, one could generate a feature at $q$%
=0 and energy of 1.2 meV as a bound state of two zone boundary T$_{1}$
phonons with opposite momenta. However, in the same way one could combine
different phonons to obtain any energy and $q$, namely this process will
produce a broad background, in contrast to what is seen in Fig {\em \ }2.
Typically, the width of multiphonon features is on the order of $\sim $%
several meV, while that of the new feature is an order of magnitude smaller
(about 0.2 meV). Next, in Fig. 4 we plot the measured linewidth of all the
phonons along (110) vs. their energy. To obtain the data in this figure, we
used well defined, symmetric phonon groups. The phonon groups were fit to a
damped harmonic oscillator function folded with the gaussian spectrometer
resolution function determined by scattering from Vanadium sample. One can
see that phonons with energy less than 1.2 meV have widths which are below
our resolution limit ($\leq $0.02 meV), while above that energy there is a
marked increase of the linewidth. In contrast, the linewidth predicted by
the SCP theory\cite{koehler3} due to phonon interaction is a smoothly
increasing function of energy\ (see Fig. 4). The abrupt increase shown by
the data strongly suggests that the feature at 1.23 meV is a distinct
excitation which the high energy phonons can decay into.

We outline here possible interpretation of the data involving{\em \ }%
localized excitations. Here, localized excitations can be point defects or
''local modes''\cite{sievers}. This approach is appealing since point
defects in solid He are predicted to add another excitation branch to the
usual phonon spectrum\cite{hetherington,guyer2,andreev}. Interpretation of
the new feature in these terms is supported by thermally activated mass \cite
{dyumin} and spin diffusion\cite{grigor'ev} experiments in bcc $^{4}$He, in
which the activation energy of point defect was found to be 1.25 meV. One
possibility is that the new observed mode is connected to a point defect
such as vacancy, interstitial, or a Frenkel pair, involving a {\em %
displacement} of atoms from their lattice sites. Several authors proposed
that point defects in quantum crystals are delocalized objects and can
behave as quasiparticles, having a finite bandwidth and well defined
dispersion relation\cite{hetherington,guyer2,andreev}. Indeed, the data for
the new feature in Fig. 2 can be fit by an expression of the type $%
E(q)=\epsilon _{0}+\hbar ^{2}q^{2}/2m^{\ast }$, with $\epsilon _{0}$=1.23
meV and $m^{\ast }\simeq $ 0.2 m$_{4}$. We cannot determine the bandwidth
directly because the new mode cannot be resolved for $q\geq $0.2 r.l.u.
There are however problems associated with this interpretation: First,
according to the data shown in Fig. 4, phonons can decay into this mode once
their energy exceeds $\epsilon _{0}$. For the L branch, the phonon with the
lowest energy(=$\epsilon _{0}$) that can decay has $q$=0.15 r.l.u. If the
mode has intrinsic dispersion, then there is a problem with momentum
conservation. Since the decay is observed, it is possible that the new mode
is intrinsically dispersionless, and the apparent dispersion seen at finite $%
q$ can be attributed to mode coupling effects\cite{wood,schober}. A
dispersionless mode is not what these models predict\cite
{hetherington,guyer2,andreev}. Second, these models assume that the energy
spectrum of point defects exists along every direction in the crystal. Our
preliminary measurements done along the [100] and [111] directions do not
show the new mode.

The other possibility is an interpretation according to the new model\cite
{gov}. \ In this model, this new excitation represents a density fluctuation
localized over two unit cells. Unlike the point defects described above,
this excitation is a {\em dynamic} density fluctuation, in which atoms
remain associated with their lattice sites. First, the model predicts
correctly the energy of this new excitation at $q$=0. Second, as fig. 3
shows, both the L and T$_{2}$ phonons become damped above the same energy of
1.2 meV. For the L phonon this happens at $q$=0.15 r.l.u., while for the T$%
_{2}$ phonon the same thing happens at $q$=0.27 r.l.u. Hence, the new
feature seems to be essentially dispersionless, as predicted\cite{gov}. 
However, one problem with this interpretation is that coupling of the new
mode with other phonons, leading to the apparent dispersion and phonon
decay, is not treated within the model.{\em \ }

Finally, model calculations done on strongly anharmonic solids indicate the
possible existence of localized modes\cite{sievers}. It would be interesting
to extend these calculations using a potential more specific to solid $^{4}$%
He, to allow a comparison with experiment.

In conclusion, we have observed a new excitation in bcc $^{4}$He. One
interpretation which we suggest views this excitation as a localized
excitation unique to a quantum crystal. The nature of the localized
excitation relies on the specific way in which atomic zero point motion is
incorporated into the theoretical models\cite
{gov,hetherington,guyer2,andreev}.

Acknowledgments: {\em \ }{\it We thank S. Hoida and M. Ayalon (Technion), S.
Raymond, F. Thomas, S. Pujol and J. Previtali (ILL) for their help with the
experiment. We have benefitted from discussions with B. F\aa k, H. R. Glyde,
A. Auerbach, B. Shapiro, and S. G. Lipson. This work was supported in part
by The Israel Science Foundation, and by the Technion Fund for Promotion of
Research.}

\bigskip

Figure Captions

{\bf Fig. 1} Intensity map of longitudinal scans along [110] showing the L
branch and an ''optic like'' excitation band. Note that the intensity scale
in the figure is logarithmic, given in neutron counts.

{\bf Fig. 2} Typical neutron scattering spectra along [110] showing the new
feature for two values of $q$. In (a), taken at $q$=0.1, the feature appears
together with the L phonon. In (b), taken at the zone origin ($q$=0), only
the new feature is seen. The lines in the figure are gaussian, and serve as
guide to the eye.

{\bf Fig. 3} Dispersion curves along [110]. Solid symbols - data for the 3
phonon branches. Open symbols-the new feature. Solid lines represent the SCP
theory \cite{glyde2}. The additional predictions of the new model are shown
as dash-dotted line (The prediction of the new model for the L and T$_{2}$
branches are the same as of the SCP theory).

{\bf Fig. 4} Experimental phonon linewidth vs. phonon energy for all 3
branches along [110] showing the abrupt increase near 1.2 meV. Symbols are
experimental data. Dash-dot line - SCP theory for the L phonon (ref.\cite
{koehler3}). The dashed line is a guide to the eye.

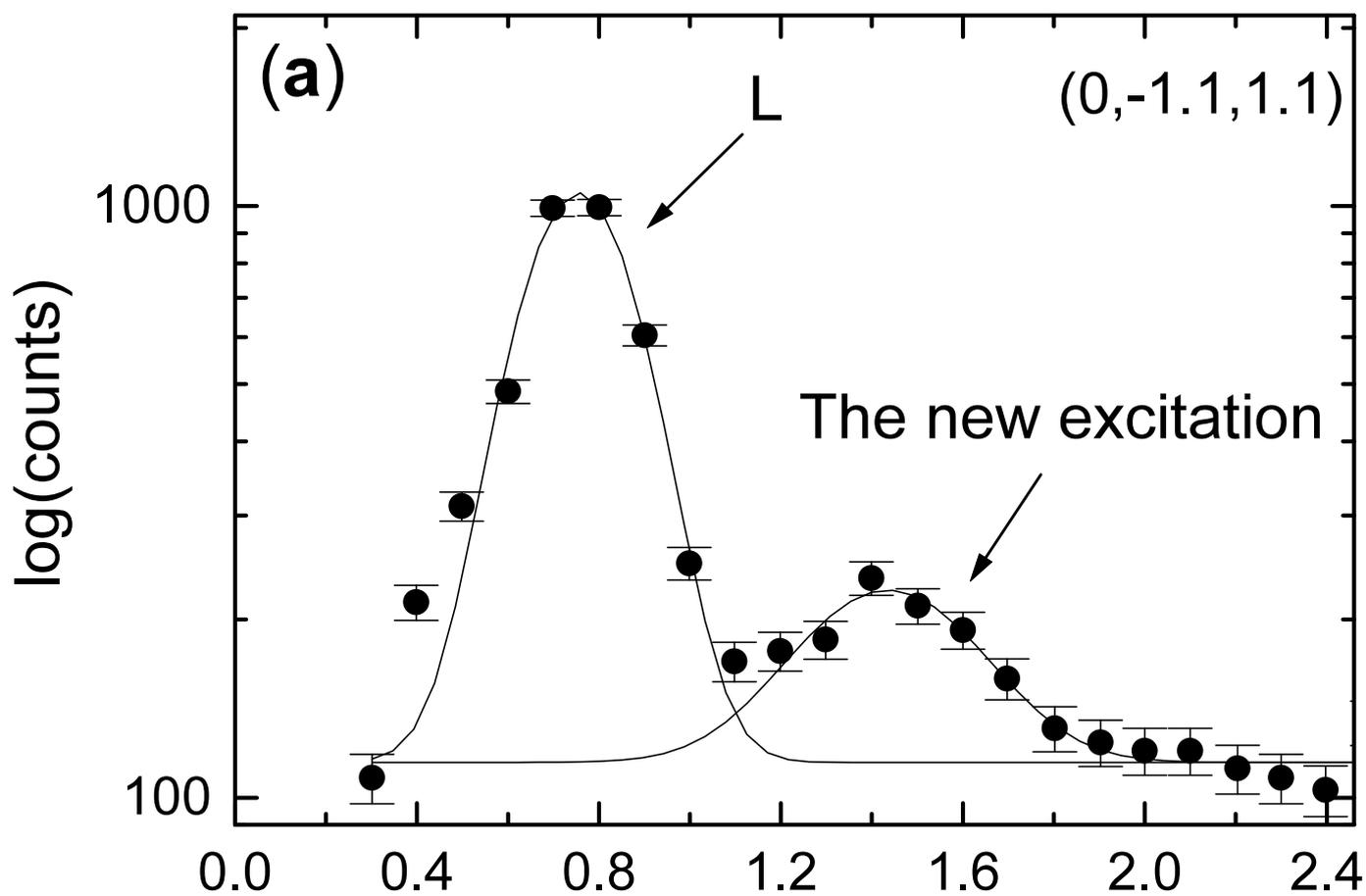
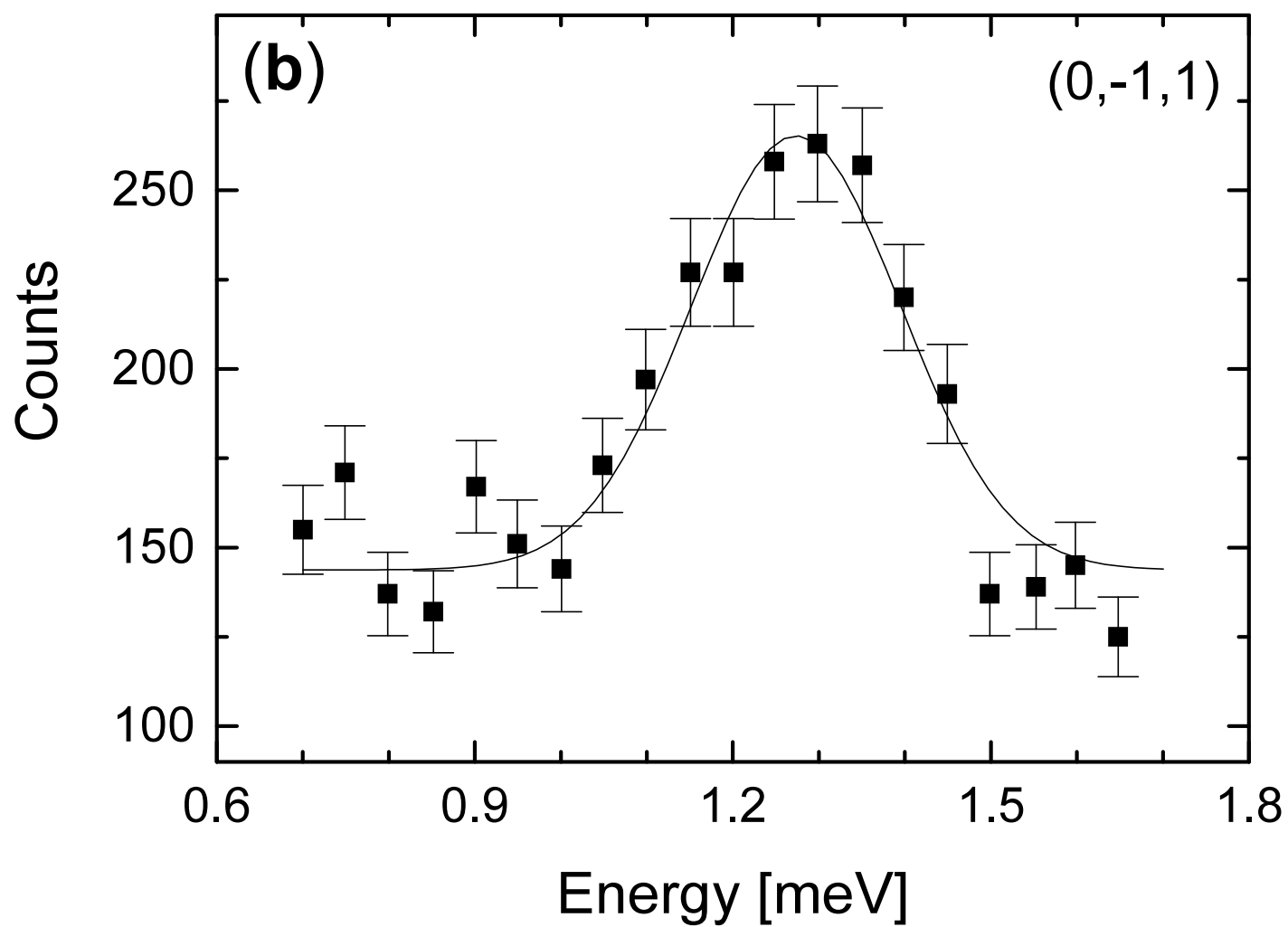
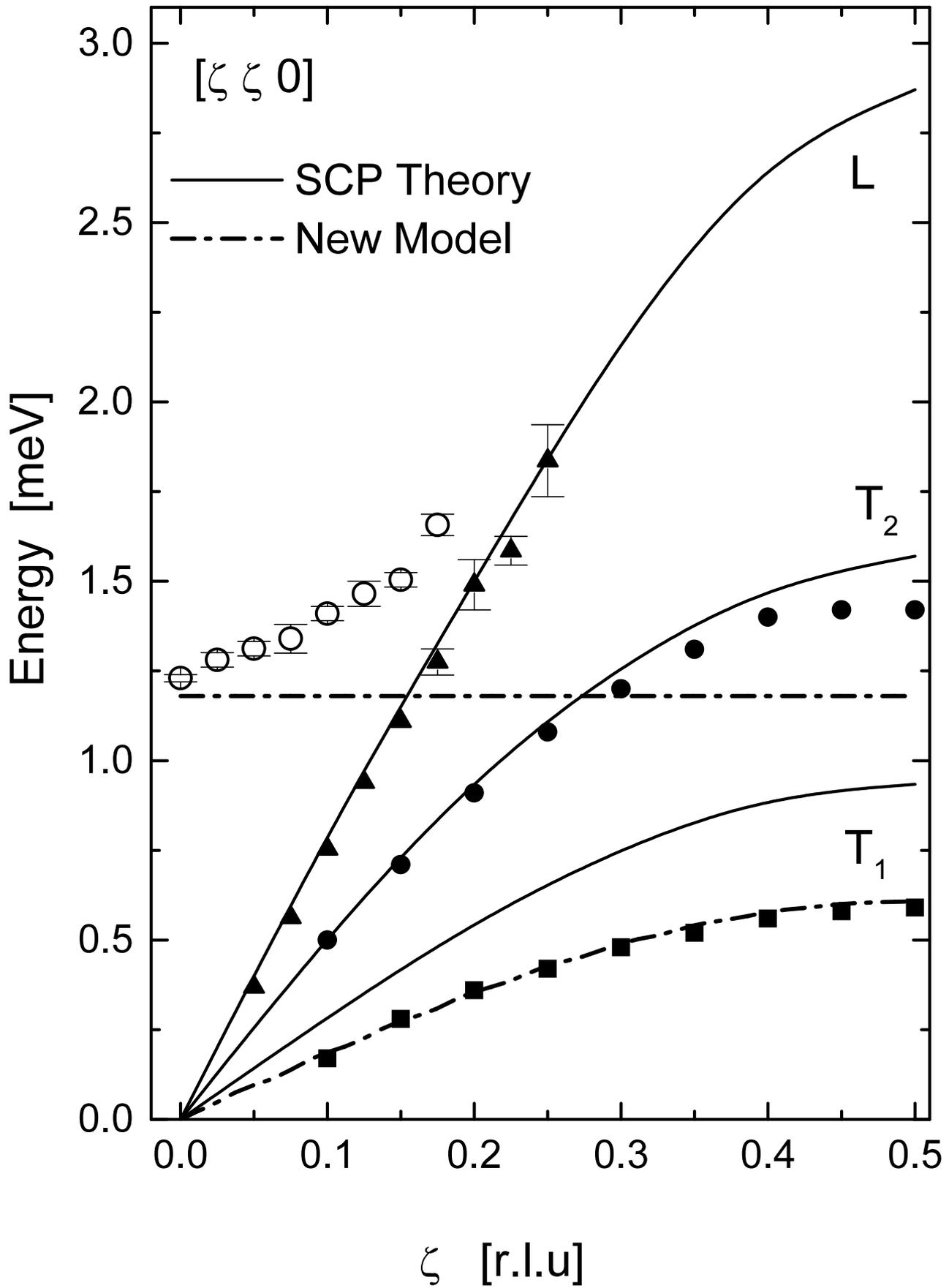
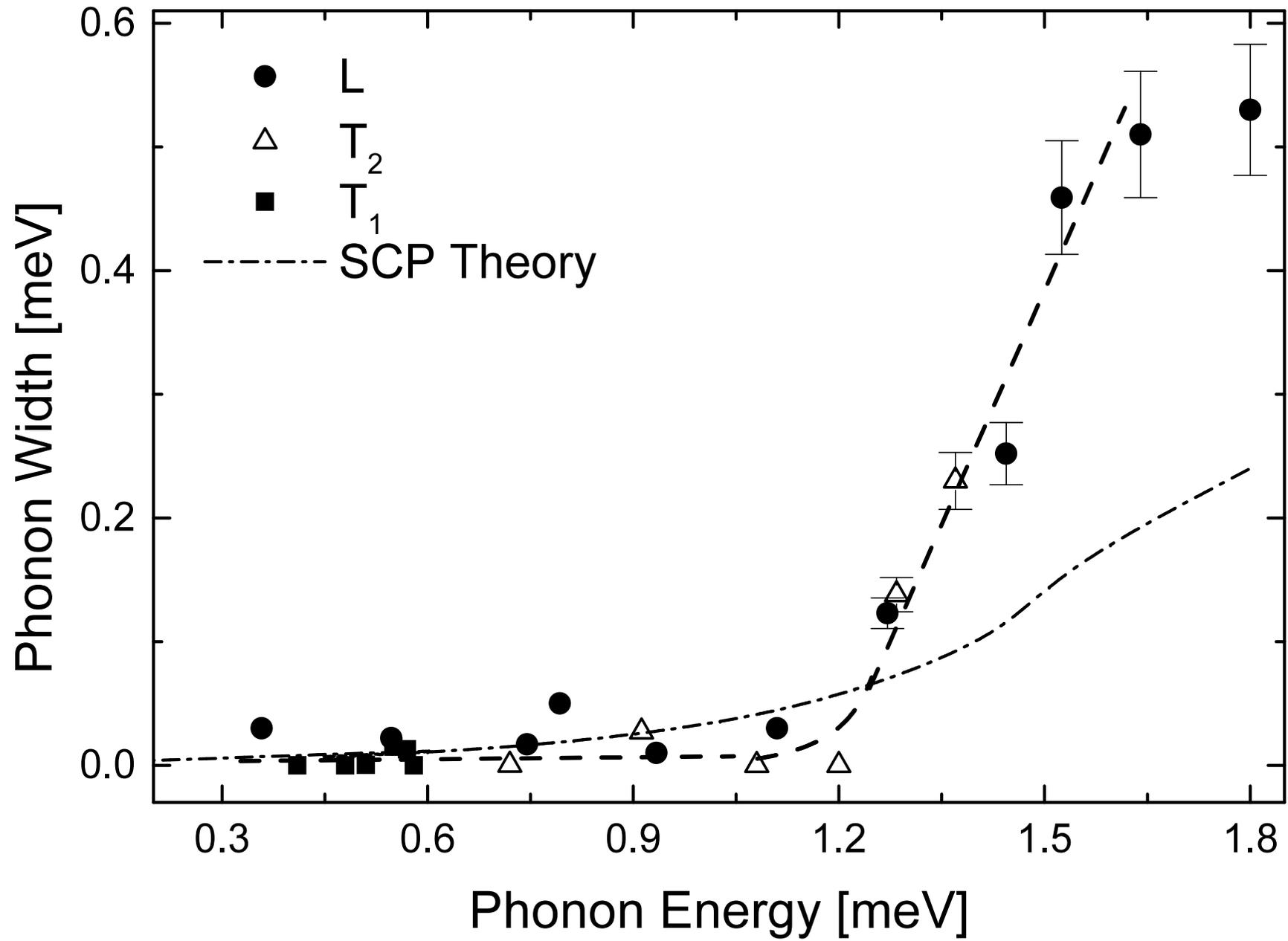
\end{document}